\documentclass[prb,twocolumn,showpacs,preprintnumbers,amsmath,amssymb]{revtex4}
\usepackage{graphicx}
\usepackage{dcolumn}
\usepackage{bm}
\begin{document}

\title{Two  Impurity Anderson problem:\\ Kondo-doublets beyond the Kondo Limit.}
\author{J. Simonin}
\affiliation{Centro At\'{o}mico Bariloche, Comisi\'{o}n Nacional de Energ\'{i}a At\'{o}mica,\\
8400 S.C. de Bariloche, R\'{i}o Negro,  Argentina}
\date{marzo 2007}

\begin{abstract}
We analyze the effects of high energy configurations on the Kondo-doublet interaction between two Anderson impurities. We found that the Kondo doublet states are robust and that their coherence energy is incremented by the inclusion of high energy configurations.  Analytic expressions are obtained for the corrections near the Kondo limit. Analysis of the system in the intermediate valence regime shows that the behavior of the system can be changed from ferromagnetic to slightly antiferromagnetic by tuning the system parameters; this regime can also be used  to study the interplay between \textit{hole}-driven and \textit{electron}-driven coherence effects. 
\end{abstract}
\maketitle
\section{Introduction}
The Anderson Impurity problem describes the interaction between a localizaed correlated state and a set of uncorrelated orbitals. This ideal system has various physical realizations, ranging from the ``classical" magnetic impurity in a metal case  to the Kondo quantum dots nanostructures. In the Kondo regime\cite{hewson} a correlated many-body state is formed, in which a cloud of extended hole states screen the impurity magnetic moment.  When two of these ``impurities" are put close enough the Two Impurity Anderson  (TIA) problem is set. Magnetic correlations between the impurities arise, mediated by the extended states of the host. For nanostructures, these magnetic correlations play a central role in the design of miniaturized spin-based devices\cite{adatoms} made of quantum dots (QD). For the ``classical" magnetic impurity problem, the TIA is a fundamental step towards the understanding of heavy fermions systems\cite{coleman,varma2}.  Man-made nanostructures that mimics few Anderson impurity arrays have made the full understanding of the correlations present in these systems a top priority, because of their possible applications in quantum information\cite{craig,sasaki,adatoms}. 

For many years the numerical analysis of Ref.[\cite{wilkins}] has been thought to describe the physics of the two impurity system in the Kondo limit. As a result the general belief\cite{coleman, adatoms} is that magnetic correlations between impurities are generated by the Ruderman-Kittel-Kasuya-Yosida \cite{kittel} (RKKY) interaction and that the Kondo effect screens such correlations. Very recently it appeared\cite{indisw,varma2,martin,konik} theoretical evidence that some very important aspects of the problem was missed in the cited analysis. In fact, it has been found in Refs.[\cite{jsfull,jsqdqw,jscond}] that there is a series of two impurity Kondo structures that lead to the formation of a ferromagnetic singlet as the ground state of the system. 

The first of that structures corresponds to the formation of a Kondo-doublet state, a very simple state mediated by the interchange of one hole between the impurities. That interchange is maximized when the impurity spins are aligned, determining the ferromagnetic character of this stage and of the final singlet ground state. The relevance of these Kondo-doublet states is two handed, they are a \textit{constructive proof} of the Kondo-doublet interaction and also variational ground states of the spin one half ($S=1/2$) subspaces.  The final stage, which corresponds to the formation of the singlet ground state, can be interpreted as a ``composite" single-impurity Kondo-singlet, in which what is screened by the formation of the Kondo-singlet is the doublet spin, not the impurity ones, which remain ferromagnetically aligned. This is called a super-singlet in Ref.[\cite{jsfull}].  

Most of the ``Kondo" correlation energy associated with the formation of the TIA singlet is generated by the formation of the Kondo-doublets. The Kondo-doublets correlation energy has been shown to be higher than the RKKY one for most of the parameter space of the Hamiltonian. Here we analyze the behavior of the Kondo doublet states beyond the Kondo limit. The effect of high energy configurations is analyzed. As those configurations are directly connected to the vertex of the Kondo-doublets, they can be considered order zero corrections to the doublets. We found that the Kondo doublet interaction is robust, and its coherence energy is incremented by the synergy of the doublets with the new considered configurations. The interplay between the different coherence channels that appear when the high energy configurations are considered can be used to manufacture TIA systems of predetermined  characteristics.

This paper is organized as follows: In Section \ref{ham} we write down the TIA Hamiltonian and a brief description of the Kondo-doublet interaction is given. An analysis of the symmetry of the doublets is performed.  In Section \ref{model} we evaluate the odd doublet variational wave function (VWF) with the added high energy configurations. In Sections \ref{uefe} and \ref{defe} we analyze the response of the doublets in the different regimes. In Section \ref{fin} we summarize our results and a qualitative interpretation of the experimental data of Refs.[\cite{craig, adatoms}] is given. In the Appendix we discuss some aspects of the variational $J_n/N_S$ expansion method we used to analyze the system.

\section{Hamiltonian and Kondo Limit Doublets}\label{ham}

The Two Impurity Anderson  Hamiltonian is the sum of the band, hybridization ($H_V$), and impurity (located at $R_j=\pm R/2$ over the $x$-axis) terms
\begin{eqnarray}\label{hamil}
H=\sum_{k \sigma} e_{k \sigma} \ c^\dag_{k\sigma} c_{k\sigma} + \textbf{v} \sum_{j k \sigma}(e^{i\; k R_j} \ \ d^\dag_{j\sigma} c_{k\sigma}+ h.c. )\nonumber \\
 - E_d \sum_{j \sigma} d^\dag_{j\sigma} d_{j\sigma} + U \sum_j d^\dag_{j\uparrow} d_{j\uparrow}d^\dag_{j\downarrow}d_{j\downarrow}\ , \ \ \ \ \ \ \ \
\end{eqnarray}
where the fermion operator $c_{k\sigma}$($d_{j\sigma}$) acts on the conduction band $k$-state (on the impurity (or QD) at $R_j$) and $\textbf{v}=V/\sqrt{n_b}\ $ is the hybridization  divided by the square root of the number of band states. Single state energies are referred to the Fermi energy. We renormalize the vacuum (denoted by $|F\rangle$) to be the conduction band filled up to the Fermi energy ($E_F$), and we make an electron-hole transformation for band states below the Fermi level: $b^\dag_{\overline{k} \overline{\sigma}}\equiv c_{k\sigma}$.  We use in the text a ``ket" notation for the impurity configurations, the first symbols indicate the status of the left impurity (the one at $x=-R/2$) and the second ones the status of the impurity on the right, e.g. $\ | 0 ,\uparrow \rangle \equiv d^\dag_{R\uparrow}|F\rangle$, $|\!\! \downarrow , \uparrow \rangle \equiv d^\dag_{L\downarrow} d^\dag_{R\uparrow} |F\rangle$, $|\!\!\downarrow,\downarrow \uparrow \rangle \equiv d^\dag_{L\downarrow} d^\dag_{R\downarrow} d^\dag_{R\uparrow} |F\rangle$.

\begin{figure}[h]
\includegraphics[width=\columnwidth]{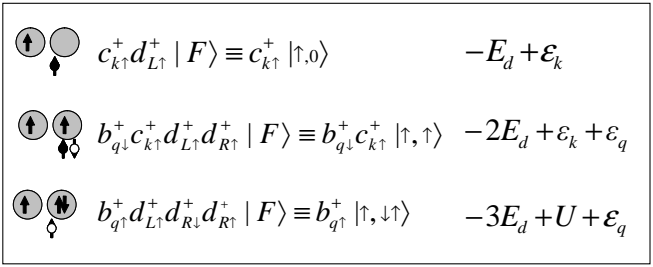}
\caption{Some configurations that are present in the Kondo-doublet and RKKY interactions; their notation and energy. The little black (white) dot corresponds to an electron (hole) excitation in the band. The gray circles represent the impurities.}\label{fig1}
\end{figure}

In the Kondo limit the impurity level is well below the Fermi energy and it can not be doubly occupied due to the strong Coulomb repulsion ($0 \ll E_d \ll U$). In this regime the two relevant  parameters of the TIA are the effective Kondo coupling $J_n=\rho_o V^2/E_d$ ($\rho_o$ being the density of band states at the Fermi level) and the interdot distance $R$. The single impurity Kondo energy is given by $\delta_K = D\ \exp{(-1/2 J_n)}$, where $D$ is the half-band width of the metal host. 

In the two impurity case, when the impurities are close enough, the higher energy stage of the Kondo nucleation is generated by the Kondo-doublet interaction. In the Kondo limit\cite{jsfull} this interaction is described by the following variational wave function 
\begin{eqnarray}\label{wodd0}
|D_{o\uparrow}\rangle= |A_\uparrow\rangle + \textbf{v} \sum_{k}\  Z_2(k) (|A_{\uparrow\downarrow k}\rangle+|A_{\uparrow\uparrow k}\rangle)\  ,
\end{eqnarray}
where the vertex state is given by
\begin{equation}\label{as0}
|A_\uparrow \rangle = (d^\dag_{R\uparrow} - d^\dag_{L\uparrow})|F\rangle= |0,\uparrow\rangle - |\!\uparrow, 0\rangle \ ,
\end{equation}
and the states
\begin{subequations}
\begin{eqnarray}\label{asmk0}
|A_{\uparrow \downarrow k} \rangle =  \ b^\dag_{k \uparrow} \ (e^{-i \textbf{k.R}/2}\  |\!\uparrow, \downarrow \rangle + e^{+i \textbf{k.R}/2}\  |\!\downarrow, \uparrow \rangle )\ , \\
\label{assk0}
|A_{\uparrow \uparrow k} \rangle =  \ b^\dag_{k \downarrow} \ (e^{-i \textbf{k.R}/2}\  |\!\uparrow, \uparrow \rangle + e^{+i \textbf{k.R}/2}\ |\!\uparrow, \uparrow \rangle )\ ,
\end{eqnarray}
\end{subequations}
are the configurations that are obtained applying $H_V$ to the vertex state. Minimizing analytically, one obtains $Z_2(k) =  1 / (\delta_o(R)+e_k)$ for the amplitude, and $E_{D_o}=-2E_d-\delta_o(R)$ for the energy of the doublet, where the odd Kondo-doublet correlation energy is given by
\begin{equation}\label{do}
\delta_{o(e)}(R) = D\ \exp{(-1/(2\pm C_h(\delta,R)) J_n)} \ ,
\end{equation}
where the minus sign corresponds to the even Kondo doublet, which is the one generated form the symmetric $( |0,\uparrow\rangle + |\!\uparrow, 0\rangle )$ vertex state. The \textit{hole coherence factor} is given by
\begin{equation}\label{ch0}
C_h(\delta,R)=  \sum_h Z_2(k_h)\cos{(\bf{k_h.R})}\ /\  \sum_h Z_2(k_h),
\end{equation}
and it is, as a function of $R$, an oscillating decaying function ( $|C_h(R)|\leq1$, of period $\simeq \lambda_F$ (the Fermi wavelength), and $C_h(0)=1$, $C_h(\infty)=0$). Its range, therefore, determines the range of the Kondo-doublet interaction, \textit{i.e.} how close the impurities must be in order to significatively interact through this mechanism. Direct ``reading" of the Kondo-doublet state Eq.(\ref{wodd0}) makes clear that the Kondo-doublet interaction is produced by the interchange of one hole between the impurities using the vertex state as a virtual bridge.

Note that this wave function, Eqs.(\ref{wodd0}) to (\ref{ch0}),  is, first, a strict  \textit{constructive proof} of the Kondo-doublet interaction, and, second, a variational ground state for the $S=1/2$ subspaces of the Hamiltonian. For this last role, it comprises the lowest energy configurations of the subspace and the simplest one, and it makes use of the strongest interaction present in the Hamiltonian, therefore we expect it to be a good variational ground state of the corresponding subspace. Furthermore, it can be taken as the first step in a variational $J_n/N_S$ expansion. In fact, we follow it two more steps in Ref.[\cite{jsfull}] in order to analyze the Kondo-doublet RKKY interplay. In the following sections we analyze the effect of two kind of high energy configurations that are directly connected with the vertex state, \textit{i.e.} they can be considered order zero corrections to the Kondo-doublets.

It is wort to discuss briefly the dependence of the Kondo-doublets in the particular structure of their vertex states. If one takes just the $ |0,\uparrow\rangle $ state as the vertex state, and the corresponding  subset of the of $|A_{\sigma_L \sigma_R k} \rangle $ configurations that are connected to it by the hybridization, ones obtains for that VWF an energy $-2 E_d -\delta_K$, \textit{i.e.} no interaction between the impurities. This can be confirmed by direct inspection of the VWF, which shows that it is the product of a Kondo singlet at the left impurity ``times" a spin up electron in the right impurity. This is not strange. If one calculates the corrections to the energy of the $|\! \uparrow , \downarrow \rangle$ and $|\! \downarrow , \uparrow \rangle$ states, up to fourth order in the hybridization $V$, one finds no $R$ dependent term for them. Instead, if one calculates for the $(|\! \uparrow , \downarrow \rangle \pm|\! \downarrow , \uparrow \rangle)$ combinations, one finds that the plus combination gains an energy $\Sigma(R)$ (it is the $S_z=0$ component of the ferromagnetic triplet, an odd state), and that the minus combination lost the same amount (it is the even antiferromagnetic singlet). $\Sigma(R)$ is half the RKKY energy.  

Therefore, the particular ``structure" of the Kondo-doublet vertex states, and that of the RKKY, is nothing more than what corresponds given the symmetries of the system. Note also that these interactions appear for any combination other than the bare ones, $(0,1)$ and $(1,0)$, but they are maximal for the proper symmetrized ones $(1,\pm1)$. This point must be taken into account if one attempts to make approximations directly over the Hamiltonian, as the Schrieffer-Wolf  transformation of Ref.[\cite{indisw}] or slave boson methods, otherwise the $V^2$ Kondo-doublet interaction will be missed. 

What is not so trivial, however, is that for the RKKY  this effect translates into a transfer of energy ($\pm\Sigma(R)$) between the combinations, whereas that for the Kondo-doublet what is transferred is the ``connectivity" of the vertex state, \textit{i.e.} its ability to interconnect the $Z_2$ configurations, the $(2\pm C_h(R))$ factor in the exponential of their correlation energy. This is a consequence of the non-perturbative nature of the Kondo-doublet interaction.

The second stage\cite{wilkins,jsfull} of the Kondo nucleation corresponds to the formation of a ``composite" Kondo-singlet, in which the doublet states play the role that the impurity states play in the single impurity Kondo-singlet. The correlation energy gained in this last stage is given by: 
\begin{equation}\label{gk}
\gamma_K  \simeq D\ \exp{(-1/(1 \mp C_h(\gamma,R)) J_n)} \ . 
\end{equation}
Thus, the total correlation energy gained by the formation of the Kondo singlet in the two impurity case is given by $E_K(R) = \delta_D + \gamma_K \geq 2 \delta_K$, where $\delta_D=\max{(\delta_e, \delta_o)}$ is the correlation energy of the dominant doublet. The following relation holds $\delta_D(R)\geq \delta_K \geq \gamma_K(R)$. For a relatively strong ``overlap" of the impurities, \textit{i.e.} $1\geq|C_h(R)| \gg 0$, one has $\delta_D(R)\gg \delta_K \gg \gamma_K(R)$ and thus the Kondo-doublet states are formed at a much higher temperature than the super-singlet. 

An extreme case is $R=0$, for which $C_h \equiv 1$. At this point the odd combinations of the band states are decoupled from the impurities, and the odd Kondo-doublet is the ground state of the system. In the opposite limit, $R\mapsto\infty$, $C_h=0$, one has $\delta_D=\gamma_K=\delta_K$, and the ``super" singlet is just the product of two single impurity Kondo singlets, one at each impurity. 

\section{Kondo doublets beyond the Kondo limit}\label{model}

\begin{figure}[h]
\includegraphics[width=\columnwidth]{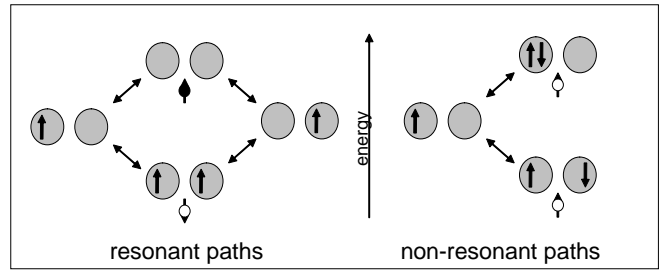}
\caption{Kondo-doublet paths, \textit{i.e.} configurations connected via the hybridization term of the Hamiltonian. The non-resonant paths are relevant to determine the full Kondo-doublet energy due to the synergy of Kondo structures.} \label{fig2}
\end{figure}

The configurations that generates the Kondo-doublet interaction, as analyzed in the previous section, are the lower energy ones of the configurations depicted in Fig.\ref{fig2}, \textit{i.e.} the ones with one electron in each QD (impurity) and one hole in the band (plus the vertex state configuration). In the resonant paths the band excitation generated when the population of one QD is modified can be reabsorbed in any of the two QDs. 

The band excitation in the non-resonant paths, instead, can only be reabsorbed at the QD at which it was generated. Nevertheless, these configurations must be included in the interaction analysis because of the strong synergy of the non-perturbative Kondo structures (Refs.[\cite{kittel}], pag.155 ,[\cite{hewson}],pag.53). In the following we analyze the effects of the high energy configurations shown in Fig.\ref{fig2}, the ones with no electrons in the QD's and the ones with a double occupied QD, on the Kondo-doublet states. These configurations are directly connected to the vertex state.

Therefore, rewriting Eqs.(\ref{as0}-\ref{asmk0}) for completeness, the properly symmetrized one-electron in-two-QDs configuration that acts as vertex of the odd Kondo-doublet is 
\begin{equation}\label{as}
|A_\sigma \rangle = (d^\dag_{R\sigma} - d^\dag_{L\sigma})|F\rangle= |0,\sigma\rangle - |\sigma, 0\rangle \ ,
\end{equation}
of energy $-E_d$. This configuration acts as a bridge between the other components of the doublet, and it is a ``virtual" state in the Kondo limit. Particularizing to the spin-up component of the doublet, the hybridization $H_V$ connect the $|A_\uparrow \rangle$ configuration with the following ones \begin{subequations}
\begin{eqnarray}\label{asmk}
|A_{\uparrow \downarrow k} \rangle =  \ b^\dag_{k \uparrow} \ (e^{-i \textbf{k.R}/2}\  |\!\uparrow, \downarrow \rangle + e^{+i \textbf{k.R}/2}\  |\!\downarrow, \uparrow \rangle )\ , \\
\label{assk}
|A_{\uparrow \uparrow k} \rangle =  \ b^\dag_{k \downarrow} \ (e^{-i \textbf{k.R}/2}\  |\!\uparrow, \uparrow \rangle + e^{+i \textbf{k.R}/2}\ |\!\uparrow, \uparrow \rangle )\ ,
\end{eqnarray}
\end{subequations}
of energy $-2 E_d + e_k$, by promoting an electron from below $k_F$ to the empty QD. In Refs.[\cite{jsfull, jsqdqw, jscond}] just these configurations were considered. The first one is a non-resonant configuration and the second one is the one that generates the strong ferromagnetic correlation between the impurities. This can be checked by the evaluation of their connectivity factors, \textit{i.e.} the configuration to vertex state matrix element of the hybridization, normalized by $- 2 \textbf{v}$
\begin{subequations}
\begin{eqnarray}\label{chsma}
\langle A_\uparrow|H_V|A_{\uparrow \downarrow k} \rangle &\sim& 1\ , \\
\label{chsmb}
\langle A_\uparrow|H_V|A_{\uparrow \uparrow k} \rangle &\sim& (1 + \cos{\textbf{k.R}}) \ ,
\end{eqnarray}
\end{subequations}
this last element depends on the inter-QD distance $R$, it generates correlations between the status of the two QDs.

The hybridization also connect the $|A_\uparrow \rangle$ state  with the following ones: the resonant
\begin{equation}
|q_\uparrow \rangle =  \ (- c^\dag_{q \uparrow} \ e^{-i \textbf{q.R}/2}\ |F \rangle + c^\dag_{q \uparrow}\ e^{+i \textbf{q.R}/2}\ |F \rangle ) \  
\end{equation}
configuration,  of energy $e_q$ and connectivity 
\begin{equation}
\langle A_\uparrow|H_V|q_\uparrow \rangle \sim\ (1-\cos{\textbf{q.R}})\ ,
\end{equation}
and the non-resonant 
\begin{equation}
|U_{k\uparrow} \rangle = \ b^\dag_{k \uparrow}\ ( e^{-i \textbf{k.R}/2}\ |0,\downarrow \uparrow \rangle -  e^{+i \textbf{k.R}/2}\ |\!\downarrow \uparrow,0 \rangle )\
\end{equation}
configuration, of energy $-2E_d + U + e_k$ and connectivity
\begin{equation}
\langle A_\uparrow|H_V|U_{k\uparrow} \rangle\sim 1 \ .
\end{equation}

Including all these states, the variational wave function for the study of the odd doublet is given by
\begin{eqnarray}\label{wodd}
|D_{o\uparrow}\rangle= |A_\uparrow\rangle + \textbf{v} \sum_{k}\  Z_2(k) (|A_{\uparrow\downarrow k}\rangle+|A_{\uparrow\uparrow k}\rangle)\ \nonumber \\
 +\ \textbf{v} \sum_{q}\  Z_0(q) |q_\uparrow \rangle  + \textbf{v} \sum_{k}\  Z_U(k) |U_{k\uparrow} \rangle \ ,
\end{eqnarray}
where the $k$ ($q$) sums are over hole (electron) excitations. In the following we will use the variational amplitude factors to refer to the corresponding set of configurations. Minimizing the expectation value of the odd doublet energy, we obtain
\begin{subequations}\begin{eqnarray}
Z_2(k) &=& -1 / (E_o+2E_d-e_k)\ , \\
Z_0(q) &=& -1 / (E_o-e_q)\ , \\
Z_U(k) &=& -1 / (E_o+2E_d-U-e_k)\ ,
\end{eqnarray}\end{subequations}
and
\begin{eqnarray}\label{enedo}
E_o=-E_d+\textbf{v}^2\sum_{k}(2 + \cos{k_x R})Z_2(k) \nonumber \\
+\textbf{v}^2\sum_{q}(1 - \cos{q_x R})Z_0(q) +\textbf{v}^2\sum_{k}Z_U(k) \ ,
\end{eqnarray}
where $E_o$ is the energy of the odd Kondo-doublet. Assuming  $E_o = - 2 E_d - \delta_o$, we rewrite the amplitude factors as
\begin{subequations}\begin{eqnarray}
Z_2(k) &=&  1 / (\delta_o+e_k)\ , \\
Z_0(q) &=&  1 / (2E_d+\delta_o+e_q)\ , \\
Z_U(k) &=&  1 / (U+\delta_o+e_k)\ ,
\end{eqnarray}\end{subequations}
and the equation for the doublet energy becomes a self-consistent equation for the doublet correlation energy $\delta_o$
\begin{eqnarray}\label{dog}
\frac{E_d+\delta_o}{V^2 \rho_0 }= I_h(\delta_o)(2+C_h(\delta_o)) + I_h(U+\delta_o) \nonumber \\
+ \ I_e(2E_d+\delta_o)(1-C_e(2E_d+\delta_o)) \ ,
\end{eqnarray}
where the Kondo integrals are given by  $I_x(\omega) = (1/n_b \rho_o) \sum_x 1/(\omega+e_x)$ and the \textit{hole} (\textit{electron}) coherence factors by
\begin{equation}\label{ch}
C_x(\omega,R)= \frac{1}{n_b \rho_o} \ (\ \sum_x \frac{\cos{\bf{k_x.R}}}{\omega+e_x}\ )\ /I_x(\omega).
\end{equation}
Further elaboration of Eq.(\ref{dog}) depends on the particular system and regime in study. The Kondo limit in 1, 2, and 3 dimensions was analyzed in Ref.[\cite{jsfull}], obtaining Eqs.(\ref{do}, \ref{gk}).  In the following we center our analysis in the finite $E_d , U$ effects generated by the high energy configurations, while using a simple model for the extended $k$ states. With the half-filled flat band model, of half-band width $D$, the Kondo integrals result to be
\begin{equation}\label{ik}
I_e(\omega)=I_h(\omega)=I_K(\omega)=\ln{\frac{D+\omega}{\omega}} \ .
\end{equation}
For the coherence factors, further choices must be made. A one dimensional system\cite{jsqdqw} has the advantage that decoherence comes only from the energy spread ($\simeq \delta_o$) of the involved excitations and not from the angular terms that are characteristic of higher dimensions\cite{jsfull}. In near 2D QDs systems made in semiconductor heterostructures such angular dispersion effects can be minimized by electron-focusing technics, rendering the system response close to the 1D results. Furthermore, in 1D the coherence factors can be analytically evaluated, which is very convenient for the purposes of the present study. Thus, particularizing for a 1D band system, the coherence factors are
\begin{eqnarray}\label{che}
C_{h(e)}(\omega,R) = [ \cos{(r^* \pm \omega^*)} [ \text{Ci}(r^* + \omega^*)-\text{Ci}(\omega^*)]  \nonumber  \\ 
\pm \sin{(r^* \pm \omega^*)} [ \text{Si}(r^* + \omega^*) - \text{Si}( \omega^*)] ]/ I_K(\omega) \ \ ,
\end{eqnarray}
where  $r^* = k_F R = 2 \pi R/\lambda_F$, $\omega^* = k_F R \ \omega/D$, and $\text{Ci}$ ($\text{Si}$) is the CosIntegral (SinIntegral) function. Both $C_h$ and $C_e$ tend to $\cos{(k_F R)}$ for $\omega \mapsto 0$, \textit{i.e.} when the band excitations that generate the correlation are the ones with $ k \simeq k_F$. 
\begin{figure}[h]
\includegraphics[width=\columnwidth]{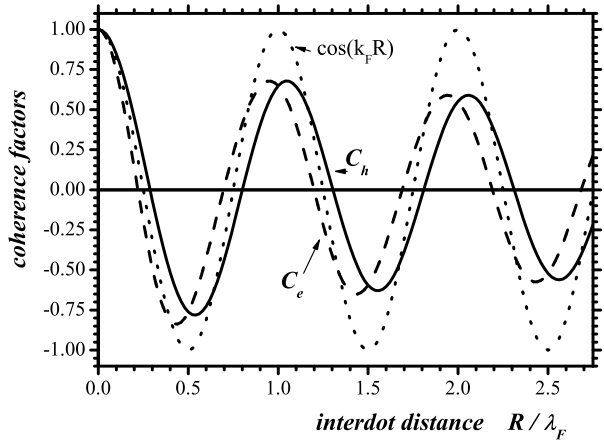}
\caption{\textit{Hole} and \textit{electron} coherence factors as a function of the inter-impurity distance, for $\omega = 0.001\ D$. In 1D their range is given by the Kondo length.}\label{fig3}
\end{figure}

In Fig.\ref{fig3} we plot the coherence factors. It can be seen that the \textit{electron} coherence factor has a period a little lower than $\lambda_F$ because it involves excitations with $k\geq k_F$, the opposite is true for the \textit{hole} coherence factor. This ``mismatch" between the coherence factors opens the possibility to design arrays with different characteristics, as we will discuss in the next sections.   

For the even Kondo-doublet, which vertex is the symmetric $ |S_\sigma \rangle = (d^\dag_{R\sigma} + d^\dag_{L\sigma}) |F\rangle  =  |0,\sigma\rangle + |\sigma, 0\rangle \ $ state, it holds the same equations but with the $Q_x\mapsto -Q_x$ change. 

The odd doublet variational wave function Eq.(\ref{wodd}) is valid for the analysis of the $0 < 2 E_d < U$ region, in which an empty impurity state (of energy $0$) is the main channel for the hybridization to act. For the $U < 2 E_d < 2 U$ region, in which the main hybridization channel is a double occupied impurity (of energy $-2E_d+U<0$),  a similar doublet generated by electron-hole symmetry arguments can be used. The vertex state of such doublet is given by
\begin{equation}\label{asu}
|U_\uparrow \rangle = |\! \downarrow \uparrow,\uparrow \rangle - |\! \uparrow, \downarrow \uparrow \rangle \ ,
\end{equation}
and the corresponding equations and properties are similar to the ones analyzed here. 
 
\subsection{Impurity Spin-Spin correlation}

To characterize the response of the TIA system in the different regimes we evaluate some observables. The most relevant is the spin-spin impurity correlation $ \langle S_L.S_R \rangle $, which clearly signals the extended nature of the Kondo doublets. The mean population of one impurity indicates the changes from Kondo to intermediate valence (or Anderson) regime. At this effect we need to evaluate the following quantities: the expectation value of the spin-spin correlation,
\begin{equation}
\langle D| S_L . S_R |D \rangle = \frac{3}{2} \ \textbf{v}^2 \ \sum_k \ \cos{k.R}\  Z_2(k)^2 \ ,
\end{equation}
of the average population of one impurity
\begin{eqnarray}
\langle D|n_R |D \rangle =\ \langle D| n_{R\uparrow}+ n_{R\downarrow}|D \rangle= \ 1 + 2\ \textbf{v}^2 \nonumber \\ \{\ \sum_k \ (2+\cos{k.R})\  Z_2(k)^2  + \sum_k \ Z_U(k)^2\   \} \ ,
\end{eqnarray}
and the norm 
\begin{eqnarray}
\langle D|D \rangle = 2 \ [\ 1 + \textbf{v}^2 \{\ \sum_k \ (2+\cos{k.R})\  Z_2(k)^2 \nonumber \\ + \sum_k \ Z_U(k)^2\ + 
 \sum_q \ (1-\cos{q.R})\ Z_0(q)^2 \  \} \ ] \ ,
\end{eqnarray}
of the doublet state. They appear two new sums, 
\begin{equation}\label{jh}
J_x(\omega)= \frac{1}{n_b \rho_o} \sum_x \frac{1}{(\omega+e_x)^2}\ \ ,
\end{equation}
and
\begin{equation}\label{dh}
D_x(\omega,R)= \frac{1}{n_b \rho_o} (\ \sum_x \frac{\cos{\bf{k_x.R}}}{(\omega+e_x)^2}\ )/J_x(\omega)\ \ ,
\end{equation}
that can be easily evaluated as derivatives with respect to $\omega$ of the previously defined $I_x$ and $C_x$. With the assumed band model it results $J_K(\omega) = D/\omega(\omega+D)$. For $D_x(\omega,R)$ it results a large expression, suffices to said that its behavior is very similar to that of $C_x(\omega,R)$; its amplitude decays a little slowly because its weighting factor ($Z_x^2$) is more biased towards the $k \simeq k_F$ excitations than the one of $C_x$ ($Z_x$). Thus the spin-spin correlation is given by 
\begin{eqnarray}\label{slsr}
\langle S_L.S_R \rangle = \langle D|S_L.S_R|D \rangle/\langle D|D \rangle = \ \ \ \  \nonumber  \\ \frac{3\ V^2 \rho_o\  D_2 \ J_2} {4 + 4\ V^2 \rho_o (J_2(2+D_2) + J_U + J_0(1-D_0))}\ \ ,
\end{eqnarray}
where $J_2=J_h(\delta_o)$, $J_U=J_h(U+\delta_o)$, and $J_0=J_e(2E_d+\delta_o)$ and the same convention is used for the $D_x(\omega)$ arguments.

\section{$U$ effects.}\label{uefe}

\begin{figure}[h]
\includegraphics[width=\columnwidth]{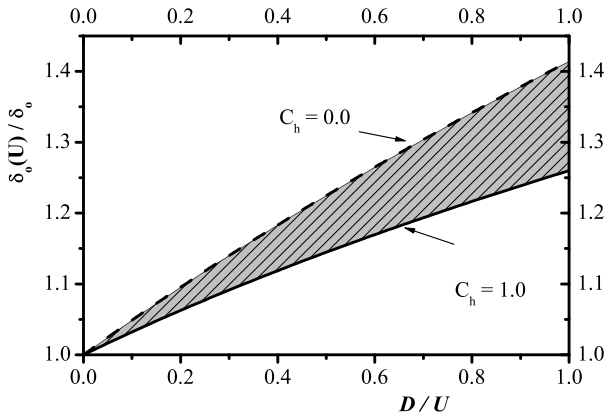}
\caption{$\delta_o(U) / \delta_o$ as a function of $D/U$. The relative change is greater the lower is $C_h$, covering the gray zone in the figure, for $C_h = 1$ to $0$. }\label{fig4}
\end{figure}

Corrections due to a finite value of $U$ are the lower ones in the Kondo doublet VWF (Eq.(\ref{wodd})), for its range of applicability.  For  $ 0 < 2 E_d < U $, it results that $Z_2 \gg Z_0 \gg Z_U$. As these corrections come from the non-resonant $Z_U$ configurations they have little dependence on $R$. Working over Eq.(\ref{dog}), and considering only the $Z_2$ and the $Z_U$ configurations effects, one obtains
\begin{equation}\label{dou}
\delta_o(U) / \delta_o = (1+D/U)^{ 1 / ( 2 + C_h(\delta_o))} \ ,
\end{equation}
which gives the incremental ratio of the Kondo doublet energy. It depends very weakly on $J_n$ and $R$, through its dependence in $C_h$. The incremental band is plotted in Fig.\ref{fig4} as a function of $D/U$.   

A remarkable aspect of this correction is its use of the synergy of Kondo structures, \textit{i.e.} the strong increment of the doublet coherence energy related to the increment of the number of available states for the jumping excitation. This effect is achieved with little effects in other properties of the doublet. At $D/U \sim 0.2$ roughly a ten percent increment of the coherence energy is obtained, whereas that the relative weight of the $Z_U$ configurations in the VWF, given by $\sum Z_U^2 / 2 \sum Z_2^2$ $ = J_h(U)/2 J_h(\delta_o)$ $\simeq \delta_o D/ 2 U^2$, is of the order of $10 ^{-4}$. Thus the spin-spin correlation generated by the Kondo doublet remains nearly unchanged.

\section{$E_d$ effects.}\label{defe}

We separate the study of $E_d$ corrections on the Kondo doublet properties in two regimes. For $E_d > D$ (Kondo regime) we use the usual effective Kondo-coupling approximation, keeping constant the $V^2/E_d$ ratio, \textit{i.e.} we study our equations for fixed values of $J_n = V^2 \rho_o / E_d$. For $D > E_d > 0$ (intermediate valence or Anderson regime) the hybridization amplitude ($V$) is keep constant, thus in this region we study the system response as a function of $J_D = V^2 \rho_o / D$. Due to the structure of Eq.(\ref{dog}), the one that determines $\delta_o$ for given $E_d,U,V,R$ values, the $U$ effect analyzed in the previous section is nearly the same multiplicative factor whatever the value of $E_d$, thus we disregard the $Z_U$ configurations in this section.

One of the effects that arise in considering a finite value for $E_d$ is that involved in the usual $E_d, D \gg \delta$ approximation that allows for the ``exponential" solutions of Eq.(\ref{dog}), Eqs.(\ref{do}). To visualize this point we first solve Eq.(\ref{dog}) as a function of $E_d$ at $ R = 0 $. Although this can be seen as an unrealistic situation for the ``classical" magnetic impurity problem, it is a very important setup for quantum dots laterally coupled to a quantum wire\cite{sasaki,konik}. At this value of $R$ one has $C_h, C_e = 1$, thus the equation for the coherence energy of the dominant odd doublet at $R=0$ results 
\begin{equation}\label{doe}
E_d+\delta_o= 3 \ V^2 \rho_0 \ln{\frac{\delta_o+D}{\delta_o}} \ \  , 
\end{equation}
\textit{i.e.} the same equation than for the single impurity problem, Ref.[\cite{hewson}, Eq.(7.17)] but with a connectivity factor of $3$. For the even doublet it is obtained
\begin{equation}\label{dee}
E_d+\delta_e=  \ V^2 \rho_0 (\ln{\frac{\delta_e+D}{\delta_e}} + 2 \ln{\frac{2E_d+\delta_e+D}{2E_d+\delta_e}} )\ . 
\end{equation}
See that the $Z_0$ configurations are decoupled from the odd doublet (at $R=0$, $C_e=1$), whereas that they make an important contribution to the even doublet, which in turn has its connectivity to the $Z_2$ configurations reduced to $1$. 
\begin{figure}[h]
\includegraphics[width=\columnwidth]{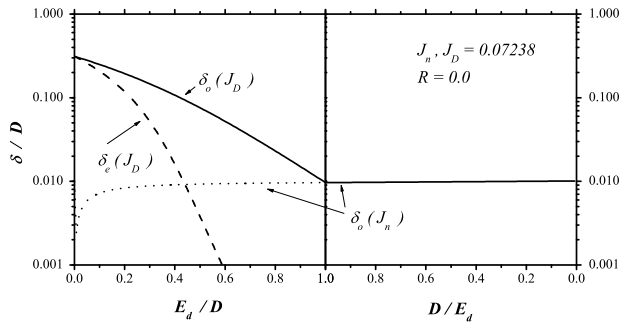}
\caption{Odd doublet coherence energy, Eq.(\ref{dog}), as a function of $E_d$ for $R=0.0$ and $J_n, J_D=0.0732$. For $E_d > D$ (Kondo regime) we keep constant the $V^2/E_d$ ratio, whereas that for $D > E_d$ (Anderson regime) the $V^2/D$ ratio is maintained constant.}\label{fig5}
\end{figure}

In Fig.\ref{fig5} we plot the odd and even doublets coherence energy as a function of $E_d$, for $R=0$ and $J_n, J_D=0.072 382$. This value of $J_n$ corresponds to a Kondo energy equal to $\delta_K=0.001\ D$ and, in the Kondo limit, to an odd doublet energy of  $\delta_3=\delta_o(R=0)=0.01\ D$. The corresponding even doublet energy is $\delta_1=\delta_e(R=0)=0.000\ 001\ D$. Lowering $E_d$ from the Kondo limit it can be seen that $\delta_o$ decreases a little from his ``exponential" expression $\delta_o=D \exp{(-1/3J_n)}$, but this is still a good approximation up to very low values of $E_d/D$, see the point line in the left panel of Fig.\ref{fig5}. This point line gives the solutions of Eq.(\ref{doe}) maintaining $V^2/E_d$ constant in the $E_d/D < 1$ region. The full line in the left panel is for a constant value of $V$, such that $J_D=J_n\ E_d/D$ remains constant, which is more in line with the experimental situation for quantum dots. The dashed line is the even doublet coherence energy, which is equal to that of the odd doublet for $E_d=0$, \textit{i.e.} when the $Z_0$ configurations have the same energy than the $Z_2$ configurations. Increasing $E_d$ makes the $Z_0$ configurations a more expensive path, and thus $\delta_e$ quickly decreases.      
\begin{figure}[h]
\includegraphics[width=\columnwidth]{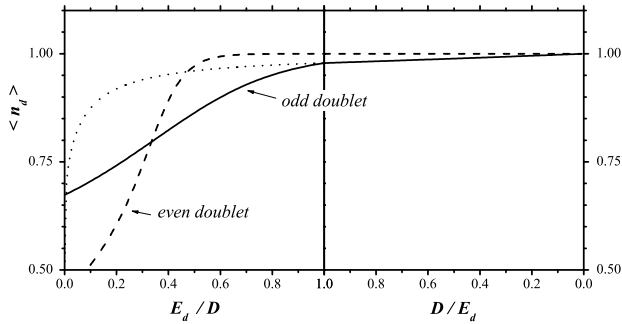}
\caption{Mean impurity occupation as a function of $E_d$, for the same parameter values than the previous figure. The main characteristic of the Kondo ($\langle n_d \rangle \simeq 1$) and Anderson (or intermediate valence) ($\langle n_d \rangle < 1$) regimes can clearly be seen. }\label{fig6}
\end{figure}

In Fig.\ref{fig6} we plot the mean population of one impurity for the same situations depicted in Fig.\ref{fig5}. These values are very similar to the ones corresponding to the single impurity case\cite{hewson}. Note that the ``Kondo coupling approximation", point line, gives higher occupation values than the ``constant-$V$" curve. This is because the former corresponds to lower values of the hybridization, such that $V^2/E_d$ remains constant. For the even doublet, as measure that $E_d$ decreases, the mean occupation remains high ($\simeq 1$) because of the reduced connectivity of its $Z_2$ configurations  until the weight of the $Z_0$ configurations becomes relevant, and then the mean impurity occupation of the even doublet becomes lower than the one of the odd doublet. 
\begin{figure}[h]
\includegraphics[width=\columnwidth]{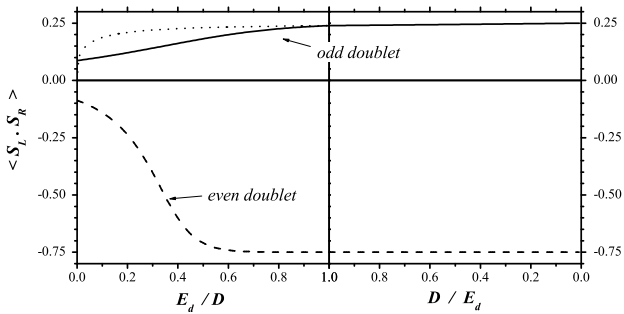}
\caption{$\langle S_L . S_R \rangle$ as a function of $E_d$, for the same parameters than the previous figures. In the Anderson regime the ferromagnetic response of the odd doublet is reduced by the increasing weight of the vertex state. For the even doublet it is also important the contribution of the $Z_0$ configurations. }\label{fig7}
\end{figure}

In Fig.\ref{fig7} we plot the Spin-Spin impurity correlation for the same situations depicted in Figs.\ref{fig5}-\ref{fig6}. This, of course, is a characteristic of the two impurity Kondo doublets that has not analogy in the single impurity case. The ferromagnetic response of the dominant odd doublet remains close to its saturation value ($\simeq 1/4$) up to low values of $E_d/D$, where the weight of the vertex state starts to be significative. For the even doublet the same effects that determine its mean impurity population curve can be traced in its Spin-Spin correlation.  

\subsection{Kondo regime}

In this region ($ E_d > D $) the usual $E_d,D \gg \delta$ approximation is a good one, and we use it to recast the solutions of Eq.(\ref{doe}) in an ``exponential" form.  Depending on the value of $J_n$ (\textit{i.e.} $V$) this is a good approximation down to even lower values of $E_d/D$, see the point lines in the left panel of Figs.\ref{fig5}-\ref{fig6}. In this regime the effect of the $Z_0$ configurations in the odd doublet is similar to that of the previously discussed $Z_U$ configurations. For the relative increment of the Kondo doublet energy one obtains
\begin{equation}\label{doee}
\delta_o(E_d) / \delta_o = (1+D/2 E_d)^{ (1-C_e(2E_d)) / ( 2 + C_h(\delta_o))} \ ,
\end{equation}
the main difference with the $Z_U$ configurations case is that the connectivity of the $Z_0$ configurations in the odd doublet depends on the \textit{electron} coherence factor. This is reflected in the numerator of the exponent in Eq.(\ref{doee}), $ 1-C_e(2E_d) $, which reduce the size of these corrections compared with those induced by the $Z_U$ configurations. Again, this increment in the coherence energy of the doublet is obtained with a very little participation of the involved configurations in the total weight of the doublet wave function, and thus other properties as the mean impurity population or the impurity Spin-Spin correlation remain nearly untouched.

\subsection{Anderson regime}

In this region ($ E_d < D $) there is a rich new structure in the Kondo doublets behavior due to the influence of the $Z_0$ configurations and their \textit{electron} driven coherence factor. The relative weight of these configurations in the doublet processes increases as measure that $E_d$ decreases, \textit{i.e.} when their energy becomes comparable to that of the $Z_2$ configurations, making them a not too expensive path for the jumping excitation. Moreover, these paths are resonant, their connectivity depending on the \textit{electron} coherence factor. As the ``periods" of the \textit{electron} and \textit{hole} coherence factor are respectively a little shorter and a little longer than $\lambda_F$, a fine tuning of the parameters  of the system, in the quantum dots heterostructures, allows for the manufacture of arrays with different characteristics.

\begin{figure}[h]
\includegraphics[width=\columnwidth]{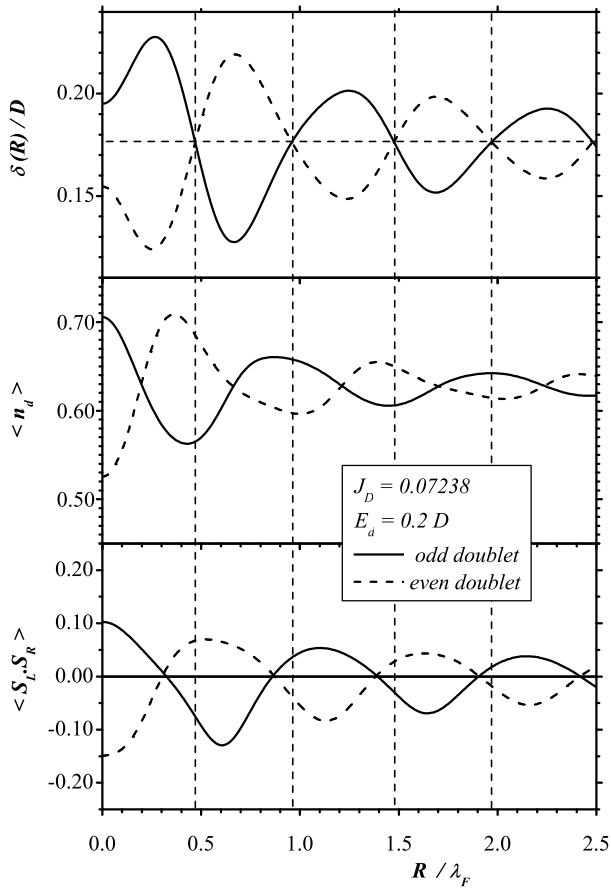}
\caption{Coherence energy, mean impurity population, and Spin-Spin correlation of the Kondo doublets as a function of the inter-impurity distance $R$, for $E_d = 0.2\ D$, and $J_D=0.0723824$. The dashed vertical lines mark the odd-even doublet transitions. }\label{fig8}
\end{figure}
In Fig.\ref{fig8} we show the Kondo doublets coherence energy, mean impurity population, and Spin-Spin correlation for a value of $E_d$ in the intermediate valence region, $E_d = 0.2\ D$, and $J_D = 0.0723824$. Due to the influence of the $Z_0$ configurations the maximum of the odd doublet energy is not at $R=0$, but at $R \simeq 0.25\ \lambda_F$, a distance that in the Kondo limit ($E_d \gg D$) is close to the first odd-even doublet transition. In the Kondo limit the coherence energy of the odd doublet decrease, as measure that $R$ increase, because the connectivity of the $Z_2$ states, given by $(2+C_h(\delta,R))$, decreases. In the present situation this fact is overcompensated by the increase of the $Z_0$ states connectivity $(1-C_e(2E_d+\delta,R))$, given that $C_e$ oscillates quickly than $C_h$. Thus the $Z_0$ configurations, decoupled from de odd doublet at $R=0$, sustain a longer first dominated region for the odd doublet and a strong maximum at $R \simeq 0.25\ \lambda_F$. These configurations also generates strong oscillations in the mean population of the impurities, because they have a null impurity population, see Fig.\ref{fig2}. And, given that the Spin-Spin correlation depends mainly in $D_h (\simeq C_h)$, Eq.(\ref{slsr}), the extended range of the odd doublet makes it to include a slightly antiferromagnetic region near the first odd-even doublet transition.   
\begin{figure}[h]
\includegraphics[width=\columnwidth]{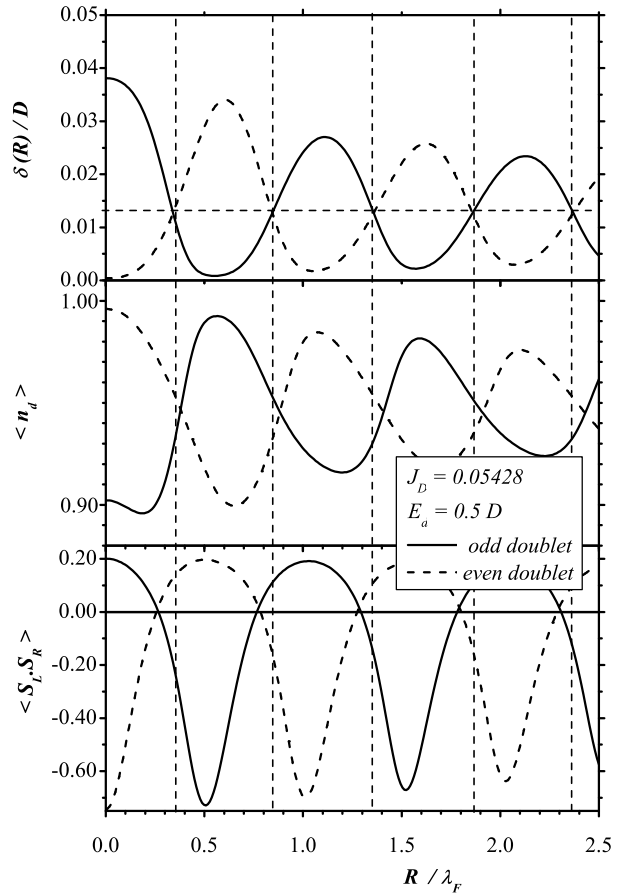}
\caption{Coherence energy, mean impurity population, and Spin-Spin correlation of the Kondo doublets as a function of the inter-impurity distance $R$, for $E_d = 0.5\ D$ and $J_D=0.05428$. At this value of $E_d$ the response of the system is already similar to that of the Kondo limit.}\label{fig9}
\end{figure}

In Fig.\ref{fig9} we show the response of the system for a greater value of $E_d$ than in previous case. For $E_d=0.5\ D$, and $J_D=0.05428$, the general behavior of the system, as a function of $R$, is already similar to that of the Kondo limit\cite{jsfull}. The mean population of the impurities is closer to one and has lower amplitude oscillations than in the previously analyzed case. The value of the ferromagnetic correlation of the dominant doublet is also closer to its saturation limit. The effect of the $Z_0$ states can be seen in the still extended first odd doublet region.  

\section{Corrections in the RKKY}

\begin{figure}[h]
\includegraphics[width=\columnwidth]{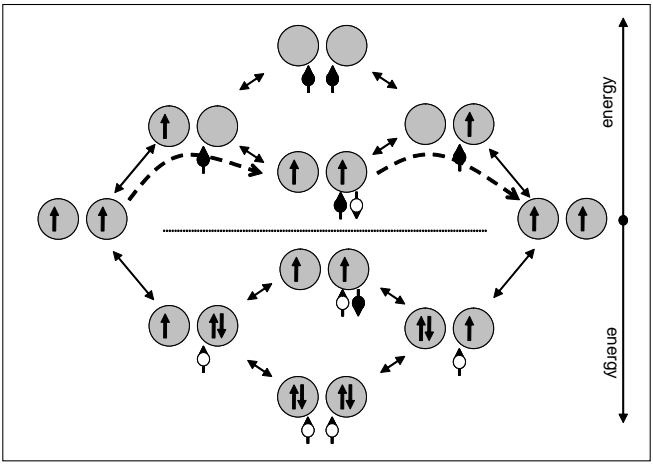}
\caption{Paths corresponding to the FM RKKY-like interactions. Fourth order paths that only contribute to single-impurity corrections are not shown. The dashed lines mark the RKKY path. }\label{fig10}
\end{figure}

The inclusion of high energy configurations in the analysis of the interactions present in the Two Impurity Anderson Hamiltonian also generates fourth order terms others than the usually discussed RKKY interaction. In Fig.\ref{fig10} we depicted all the paths that generate fourth order correlations between two aligned Anderson impurities. The dashed curved lines mark the RKKY path. The upper most path, that goes through the full empty impurity configurations, is the superexchange\cite{superex,lucio} (or double exchange) one. The lower ones, below the horizontal line at the middle of the figure, are the ``hole-electron" symmetric of the upper ones.  

Although usually ignored in applications of the RKKY, these terms has been known for a long time. They were reported in Ref.[\cite{falicov}], and carefully analyzed in Ref.[\cite{cesar}]. The main result is that the superexchange usually competes with the RKKY, \textit{i.e.} they have opposite signs, and eventually it dominates the RKKY for short distances between the impurities. These perturbative interactions are additive, there is not synergetic effects for them. More recently, they have been re obtained in Ref.[\cite{indisw}]. 

\section{Conclusions}\label{fin}

We analyze the effect of high energy configurations on the Kondo doublet interaction. We find analytic expressions for the corrections to the Kondo doublet coherence energy around the Kondo limit. We show that in that limit a relatively strong increment of the energy is produced without significatively affect others properties of the doublets. These corrections, due to the full empty and double occupied impurity configurations, are multiplicative. 

In the intermediate valence region we find that, as measure that full empty impurity configurations becomes energetically accessible, there is an interesting interplay between \textit{hole} and \textit{electron} driven coherence effects. The maximum of the odd doublet is shifted from $R=0$ towards $R\simeq \lambda_F/4$ and the first $R$-region dominated by the odd doublet is significatively increased, making a slightly  antiferromagnetic region accessible.  

As pointed out in Ref.[\cite{varma2}], the experimental results of Craig \textit{et al}\cite{craig} can be interpreted as the resonances corresponding to the odd and even Kondo doublet states. But accordingly with our previous results\cite{jsfull}, we can not rule out that the lower resonance corresponds instead to the formation of the two impurity Kondo super-singlet. Our results for the odd-even Kondo doublet transitions as a function of the inter-impurity distance are compatible with the data of Wahl \textit{et al}\cite{adatoms} for Co adatoms on a cupper surface, a detailed analysis will be present elsewhere. We must pointed out that both the authors of Ref.[\cite{craig}] and Ref.[\cite{adatoms}] interpreted their results in terms of the RKKY exchange interaction.

Therefore, we have shown that the Kondo doublet interaction is robust beyond the Kondo limit. The interplay between the different coherence channels can be used to manufacture two impurity systems of predetermined  characteristics.
\begin{acknowledgments}
I thanks the CONICET (Argentina) for partial financial support.
\end{acknowledgments}
\appendix*
\section{On the $J_n/N_S$ Expansion}
To analyze the Two Impurity Anderson system we use an extension of the ``variational $1/N_S$ expansion", Ref.[\cite{hewson}] pag.223 (hereafter HW), to the few impurity case\cite{lucio,jsfull}. This method is based in the  Kondo singlet VWF designed by Varma and Yafet\cite{varma}. It was later used by Gunnarsson and Sch\"{o}nhammer\cite{gunna} to successfully explain the spectra of rare earth compounds, and has become known as the Variational $1/N_S$ Expansion,  where $N_S$ is the degeneracy of the localized orbital. Although detailed in Hewson's book, we point here to some aspects of the method not covered there.

First, the connection of the method with the Varma and Yafet singlet, see HW-Eq.(8.44), that is the VF singlet. VF work is not cited in HW. 

Second, the expansion parameter is $J_n/N_S$, not just $1/N_S$. See HW-Fig. 8.20. Each new generation of states that are included in the variational wavefunction is comprised of the new configurations that are obtained from the previous one by the application of $H_V$. Therefore, calling generation zero to the starting $|F\rangle$ state (taken also as the energy reference), odd generations have one electron in the impurity and their energies start from $-E_d$ , and even generations have zero (or two) electrons in the impurity, and their energies start from zero (or $-2E_d+U>0$). In the Kondo limit the even generations play the role of nearly virtual states connecting the odd generations states (their total weight in the VWF is much lower than that of the odd generations). Thus, a factor $\rho_o V^2/E_d = J_n$ appears every two generations. Typical values of $J_n$ are $0.1$, for $\delta_K \simeq 0.01 D$, to  $0.05$, for which $\delta_K \simeq 0.0001 D$. The results of this method are exact in the $J_n/N_S \rightarrow 0$ limit. 

Third, range of application. Taking into account the VF VWF, and the first correction to it, the approximate ground state for the single impurity Anderson system is  
\begin{equation}\label{sk}
|S_K \rangle = |F\rangle + \sum_{k \sigma} Z_k  \ b^\dag_{k \sigma}  |\overline{\sigma}\rangle + \sum_{k q \sigma} Y_{k q}  \ b^\dag_{k \sigma} c^\dag_{q \overline{\sigma}} |F\rangle \ ,
\end{equation}
the variational amplitudes result to be $Z_k = \textbf{v} /(\delta_K+e_k)$, and $Y_{k q}= \textbf{v} \ Z_k/(-E_S+e_k+e_q)$. The energy of the Kondo singlet is given by 
\begin{equation}\label{es}
E_S=-E_d  -\delta_K + E_I \ ,
\end{equation}
where the Kondo energy $\delta_K$ comes from the resonance between the $Z_k$ configurations, which use the Fermi sea configuration as a nearly virtual bridge. $E_I$ is the single-body impurity correction, equal to $-J/2$ for the level of approximation used in Eq.(\ref{sk}). It comes from the interplay between each $Z_{k_o}$ configuration and their derived $Y_{k_o q}$ configurations. Therefore Eq.(\ref{es}) ilustrates the versatility of this variational method: it gives both the non-perturbative\cite{kittel} Kondo term as well as the standard perturbative ones. At this level, if one wants to compute corrections in $U$, one must also include the configurations with two electrons in the impurity, $\ b^\dag_{k \downarrow}b^\dag_{p \uparrow}  |\! \uparrow \downarrow\rangle $, that are obtained from the $Z_k$ states, these configurations are not shown in HW-Fig. 8.20. In this way the VWF can be expanded to the desired level of accuracy, depending on the situation to analyze, see for example the works by Gunnarsson and Sch\"{o}nhammer. If few generations are needed, results can be obtained in a nearly  analytical way. In other cases, the variational amplitudes can be determined by numerical methods.

The picture above is for the ``usual" Kondo regime ($ U \gg E_d \gg 0$), in which the main ``virtual" configurations are the ones with zero electron in the impurity, and the ones with two electrons are relegated to a secondary role because of their energy cost ($-2E_d+U \gg 0 $). For the $ E_d < U < 2E_d $ region, in which the single occupied impurity is still the lower energy configuration, but the double occupied one is the main ``intermediate" state ($-E_d < -2E_d+U <0 $), an electron $\leftrightarrow$ hole mapping must be done. In this region the VF singlet reads 
\begin{equation}\label{sku}
|S_K^U \rangle = |\! \uparrow \downarrow\rangle + \sum_{q \sigma} Z_q  \ c^\dag_{q \sigma}  |\overline{\sigma}\rangle  \  ,
\end{equation}
and it can be expanded to any desired level of accuracy in the same way than in the previously discussed region.

Therefore, this method can be applied \textit{almost anywhere} in the space of parameters. The $U\equiv2E_d$ can not be directly accessed, but it can be approached from above and below. This method can be easily extended to the few impurities cases\cite{lucio,jsfull}.  
\bibliography{beyond}

\end{document}